\documentclass[prd,twocolumn,nofootinbib,draft]{revtex4}

\newcommand{\newc}{\newcommand}
\newc\eg{{\it {e.g.}}}  \newc\etal{{\it {et al.}}} \newc\ie{{\it i.e.}}
\newc\etc{{\it {etc}}}  \newc\ibid{{\it {ibid}}}
\newc\vs{{\it {vs.}}}

\newcommand\lsim{\mathrel{\rlap{\lower4pt\hbox{\hskip1pt$\sim$}}
   \raise0.2pt\hbox{$<$}}}
\newcommand\gsim{\mathrel{\rlap{\lower4pt\hbox{\hskip1pt$\sim$}}
   \raise0.2pt\hbox{$>$}}}

\newcommand\axino{\tilde{a}}        \newcommand\maxino{m_{\axino}}

\newcommand\mchi{m_{\chi}}              

\newcommand\treh{T_{R}} 

\newc{\abund}{\Omega h^2}
\newc{\omegachi}{\Omega_\chi}      \newc{\abundchi}{\Omega_\chi h^2}
\newc{\omegadm}{\Omega_{\rm DM}}   \newc{\abunddm}{\Omega_{\rm DM} h^2} 
\newc{\abundcdm}{\Omega_{{\rm CDM}} h^2}
\newc{\omegab}{\Omega_{b}}	\newc{\abundb}{\Omega_{b} h^2}

\newc{\rhocrit}{\rho_{crit}}
\newc{\rhochi}{\rho_{\chi}}

\newcommand\tev{\,\mbox{TeV}}
\newcommand\gev{\,\mbox{GeV}}
\newcommand\mev{\,\mbox{MeV}}
\newcommand\kev{\,\mbox{keV}}

\newc\snu{{\widetilde \nu}}
\newc\snue{{\widetilde \nu}_e}      \newc\msnue{m_{\snue}}
\newc\snutau{{\widetilde \nu}_\tau}      \newc\msnutau{m_{\snutau}}

\newcommand\squark{\widetilde q}

\newc\supl{{\widetilde u}_L}      \newc\msupl{m_{\supl}}
\newc\supr{{\widetilde u}_R}      \newc\msupr{m_{\supr}}
\newc\sdl{{\widetilde d}_L}      \newc\msdl{m_{\sdl}}
\newc\sdr{{\widetilde d}_R}      \newc\msdr{m_{\sdr}}

\newc{\ra}{\rightarrow}
\newc{\beq}{\begin{equation}}
\newc{\eeq}{\end{equation}}
\newc{\bea}{\begin{eqnarray}}
\newc{\eea}{\end{eqnarray}}

\begin{document}
 \preprint{ 
}

\title{Axino dark matter from $Q$-balls in Affleck-Dine baryogenesis\\
  and the {\boldmath $\omegab - \Omega_{DM}$} coincidence problem}

\author{Leszek Roszkowski$^{1}$ 
and Osamu Seto$^{2}$ 
\footnote{Present address: 
Instituto de F\'{i}sica Te\'{o}rica, Universidad Aut\'{o}noma de Madrid,
Cantoblanco, Madrid 28049, Spain.}
 }

\affiliation{(1) Department of Physics and Astronomy, University of Sheffield,
Sheffield, UK,\\
(2) Department of Physics and Astronomy, University of Sussex, 
 Brighton BN1 9QJ, UK}


\begin{abstract}
We show that the $\omegab - \Omega_{DM}$ coincidence can naturally be
explained in a framework where axino is cold dark matter which is
predominantly produced in nonthermal processes involving decays of
$Q$-balls formed in Affleck-Dine baryogenesis.  In this approach, the
similarity of $\omegab$ and $\Omega_{DM}$ is a direct consequence of
the (sub-)GeV scale of the mass of the axino, while the reheating
temperature $\treh$ must be low, some $10^2\gev$, or less.
\end{abstract}


\date{\today}
\maketitle

\noindent
{\bf 1. Introduction.}\hspace*{0.3cm}

The origin of nonbaryonic cold dark matter (DM) and of baryon
asymmetry in the Universe are among the longest lasting puzzles in
cosmology as well as in particle physics today.  In particular, the
question of why the observed values of baryon density $\omegab$ and of
dark matter $\Omega_{DM}$ are so close to each other,
$\Omega_{DM}/\omegab=5.65\pm0.58$~\cite{wmap3yr}, remains a mystery.

A standard paradigm is that the nonbaryonic cold dark matter is made
up of some weakly interacting massive particle (WIMP) which freezes
out of thermal equilibrium in the early Universe.  Perhaps the most
popular WIMP candidate is the lightest neutralino of the Minimal
Supersymmetric Standard Model (MSSM) as the lightest supersymmetric
particle (LSP). It remains stable due to the conservation of $R$-parity. This
economical scenario does not, however, explain the proximity of
$\omegab$ and $\Omega_{DM}$.

The same is generally true for conventional mechanisms of baryogenesis or
leptogenesis. This may indicate that the observed
baryon-to-DM density ratio is just a pure accident, or else a result
of some underlying, and as yet unknown, more fundamental theory.
An alternative approach is to try to identify a
physical mechanism which would simultaneously produce both baryon
asymmetry and DM in the proportions consistent with
observations. It is clear that this basically necessitates abandoning
standard paradigms for producing both types of species in the
Universe. This may be one important lesson to learn from these considerations.

A number of attempts at explaining baryon-to-DM ratio  have been
suggested in the literature. 
For instance, recently a right handed sneutrino~\cite{Hooper:2004dc}
and a sneutrino condensate as an AD field~\cite{mcd06-rhsneutrino}
have been proposed.

A few years ago, Enqvist and McDonald (EMD)
proposed~\cite{emd98-bbb,emd98-decayqballs} an attractive solution
based on a variant of Affleck-Dine (AD) baryogenesis~\cite{ad84}. In
that scenario, an AD
condensate forms during inflation and develops a large vacuum
expectation value (VEV) along a $D$-flat direction in the MSSM.
$D$-flat directions are configurations of scalar fields for which the
$D$-part of the potential vanishes.
Prevalent in theories with many scalar
fields like the MSSM, they are of much interest to cosmology~\cite{em02-pr}.

In the standard AD scenario, after the end of inflation, the scalar
field condensate slowly rolls towards the origin and, after a few
dozens of coherent oscillations, produces a nonzero baryon number in
presence of baryon number-violating couplings of the fields making up
the flat direction. Originally, Kusenko and Shaposhnikov argued that
the AD condensate can instead fragment into nontopological solitons
called $Q$-balls~\cite{ks97-qball}.  If their baryonic charge is
large enough, as in models with gauge mediated SUSY breaking, $Q$-balls
remain effectively stable until today, and contribute to the DM
density, despite severe astrophysical 
constraints~\cite{kls05}. On the other hand,
EMD demonstrated that, under nontrivial but natural
conditions (that we summarize below), in a large class of
supergravity (SUGRA) models with 
gravity mediated SUSY breaking (GRMSB)
$Q$-balls
subsequently decay into baryonic matter and neutralino WIMPs assumed
to be the LSP~\cite{emd98-bbb}.

In the EMD scenario, the baryon-to-DM ratio can easily be
estimated to be in the right ballpark, as we shall see below.  This
otherwise attractive framework suffers, however, from a serious
problem: neutralino production in $Q$-ball decays is in fact too
efficient, and density $\omegachi$ can only agree with
observations for low neutralino mass $\mchi\sim1\gev$, well below LEP
limits~\cite{pdg06}. Moreover, this puts into a potential jeopardy the
AD mechanism in a large class of GRMSB supergravity models.

In this Letter, we suggest a way out from the above problems of the
EMD scenario which at the same time preserves its successful features,
in particular, an explanation of the $\omegab/\omegadm$ ratio. We
propose that the DM is not made up of the neutralino but
instead of an axino, a superpartner of the axion. The axino is a
neutral Majorana, chiral fermion. It  arises in SUSY models
incorporating a Peccei-Quinn solution to the strong CP problem in
QCD. Unlike for the neutralino or gravitino, its mass is strongly
model dependent and can be much smaller than the (gravity mediated)
SUSY breaking scale~\cite{axinomass,axinomass-my,axinomass-gy}. 
Similarly to the axion, its
interactions are suppressed by the PQ scale $f_a\simeq10^{11}$~GeV,
well below the sensitivity of LEP.  The axino has a number of
properties which make it a promising candidate for cold dark
matter~\cite{ckr,ckkr}.  Earlier papers considered warm axino
relics~\cite{axinomass-gy,AxinoWDM}.  
As we will show below, axinos are naturally
produced at low temperatures of a few$\,\gev$, consistent with the
$Q$-ball scenario of EMD but still before the period of Big Bang
Nucleosynthesis (BBN).

\vspace{0.15cm}
\noindent
{\bf 2. The Enqvist-McDonald scenario.}\hspace*{0.3cm}

We now briefly present the main features of the EMD variant of
AD baryogenesis. It is assumed that the AD field
$\phi$ is a $D$-flat direction in the MSSM. Its potential is,
in general, lifted by soft supersymmetric (SUSY) breaking terms and
nonrenormalizable terms~\cite{Ng,DineRandallThomas}.

The potential of the AD field, including inflaton-induced terms, reads 
\begin{eqnarray}
\label{ADPotential}
V(\phi) &\simeq&
\left\{ (m_{\phi}^2- c_1 H^2)
\left[1+K\ln\left(\frac{|\phi|^2}{\Lambda^2}\right)\right]\right\} |\phi|^2
\nonumber 
\\ &&+ 
\left[\left(c_2 H+ A m_{3/2}\right)\frac{\lambda\phi^n}{nM^{n-3}}+
{\rm H.c. } \right]
 +\lambda^2\frac{|\phi|^{2n-2}}{M^{2n-6}} , \nonumber \\ 
 &&
\end{eqnarray}
where $m_{\phi}$ is the soft SUSY breaking mass for the AD field and a
radiative correction is given by $K\ln|\phi|^2$. A flat direction
dependent constant, $K$, takes values from $-0.01$ to
$-0.1$~\cite{QballEnqvist,ejm00-flatpot}.  
$\Lambda$ denotes a renormalization scale and
$-c_1 H^2$, with $c_1\sim1$, is the negative mass-squared term induced by 
the energy density of the inflaton~\cite{DineRandallThomas}.  
Terms proportional to $A$ and $c_2$
are the trilinear terms from low energy SUSY breaking and those
induced by the inflaton, respectively, while $m_{3/2}$ denotes the
gravitino mass.  The nonrenormalizable terms in
Eq.~(\ref{ADPotential}) come from the superpotential $W =
\lambda/nM^{n-3}\phi^n$,
where $\lambda$ is the Yukawa coupling and $M$ is some large scale
acting as a cut-off.  In SUGRA, it is natural to assume 
$M=M_P \simeq 2.4 \times 10^{18}\gev$ which is the reduced Planck mass.

Since during inflation the
Hubble parameter $H\gg m_{\phi}\sim m_{3/2}$, the AD field
settles down at the minimum of the potential~(\ref{ADPotential})
which is given by 
\begin{equation}
|\phi| \simeq \left(\sqrt{\frac{c_1}{n-1}}\frac{H M_P^{n-3}}{\lambda}\right)^{1/(n-2)}
 \simeq \left(\frac{H M_P^{n-3}}{\lambda}\right)^{1/(n-2)}.
\label{PhiInfDom}
\end{equation}
It is clear that the AD field can naturally develop a very large VEV,
which is possible in nonminimal K\"ahler
potentials~\cite{DineRandallThomas}, or if large enough trilinear
term $A$ is induced by the inflaton~\cite{Kasuya:2006wf}.

We have neglected in Eq.~(\ref{ADPotential})  thermal mass terms
$h^2T^2|\phi|^2$, where $h$ denotes couplings of the AD field to other
particles~\cite{ThermalMass}. They would play a role if the AD field
VEV were relatively small. 
We have also neglected two loop thermal effects due to the
running of gauge coupling which generate a
term $\alpha T^4\ln\left(|\phi|^2/T^2\right)$, where
$|\alpha|=\mathcal{O}(10^{-2})$~\cite{TwoLoop}.  They will not be
important below.

As $H$ decreases, the AD field traces the
instantaneous minimum after inflation, begins to oscillate when
$H_{\rm{osc}}^2 \simeq m_{\phi}^2$ and, after a few dozen turns,
produces a nonzero baryon number and then fragments into
$Q$-balls.


The baryon number density for the AD field $\phi$ is given by $ n_b= i
q(\dot{\phi}^*\phi-\phi^*\dot{\phi}) $ where $q$ is the baryonic
charge for the AD field.  By using the equation of motion of the AD
field, the charge density can be rewritten as
\begin{equation}
n_b(t) \simeq \frac{1}{a(t)^3}\int^t dt' a(t')^3 \frac{2q m_{3/2}}{M_P^{n-3}}
{\rm Im} (A \phi^n) ,
\end{equation}
with $a(t)$ being the scale factor.  When the AD field starts to
oscillate around the origin, the baryon number density
 is induced by the relative phase between
$A$ and $c_2$. With the entropy density after reheat $ s = 4\pi^2g_*T^3/90 $,
we can express the baryon asymmetry as 
\begin{eqnarray}
\frac{n_b}{s}
 = \left.\frac{\treh n_b}{4M_P^2H^2}\right|_{t_{\rm{osc}}}
 \simeq  \frac{q|A|m_{3/2}}{2}\frac{\treh |\phi_{\rm{osc}}|^n}
 {H_{\rm{osc}}^3 M_P^{n-1}} \sin\delta.
\label{b-sRatio}
\end{eqnarray}
Here, $t_{\rm{osc}}$ denotes the time of the start of the oscillation, and
$\sin\delta$ is the effective CP phase.

From now on, we consider the case of $n=6$ because a promising AD
field for our scenario, a $\bar{u}\bar{d}\bar{d}$ direction belongs to
this class.  Let us
first evaluate $n_b/s$. The baryon asymmetry, Eq.~(\ref{b-sRatio}),
for the relevant case is estimated as
\begin{eqnarray}
\frac{n_b}{s}
\simeq && \frac{q|A|\sin\delta}{2\lambda^{3/2}} 
 \frac{m_{3/2}\treh}{m_{\phi}^{3/2}M_P^{1/2}}
\nonumber \\
\simeq && 1\times 10^{-10} \frac{q|A|\sin\delta}{\lambda^{3/2}} 
 \left(\frac{m_{3/2}}{100\rm{GeV}}\right)
 \left(\frac{10^3 \rm{GeV}}{m_{\phi}}\right)^{3/2} \nonumber \\
 && \times \left(\frac{\treh}{100 \rm{GeV}}\right)
\label{BaryonAsymNoEarly}
\end{eqnarray}
which is of the right order. (The previously made assumption
$M\sim M_P$ is crucial, for otherwise $Q$-balls would decay too early or would
eveporate.) The low reheat temperature $\treh$
after inflation is required to explain the appropriate baryon
asymmetry.  In this case, the AD condensate fragments into
$Q$-balls~\cite{QballEnqvist,ejm00-flatpot}.

The growth of perturbations of the AD field and its subsequent
fragmentation into $Q$-balls crucially depends on the logarithmic
correction to the $\phi^2$ mass term in $V(\phi)$,
Eq.~(\ref{ADPotential}). An essential requirement is that $V(\phi)$ is
flatter than quadratic, or that $K<0$~\cite{ejm00-flatpot}. This can
be achieved in SUGRA models 
with a nonminimal K\"ahler potential~\cite{QballEnqvist}.

In order to discuss the evolution of $Q$-balls, first we briefly
summarize their relevant properties in GRMSB 
models.  The radius of a $Q$-ball, $R$, is estimated as $R^2 \simeq
2/(|K|m_{\phi}^2)$~\cite{QballEnqvist}.  The charge is roughly given
by
$Q \simeq \frac{4}{3}\pi R^3 n_{b}(t_i)
 \simeq \frac{4}{3}\pi R^3 \left(H_i/H_{\rm{osc}}\right)^2 
 n_{b}|_{t_{\rm osc}},$
where the suffix $i$ represents the time when the spatial
imhomogeneity becomes nonlinear, 
which can be evaluated as~\cite{kk00-fB}
\begin{eqnarray}
Q &\sim& 6\times 10^{-3} \frac{2q|A|\sin\delta}{\lambda^{3/2}}
\frac{m_{3/2}M_P^{3/2}}{m_{\phi}^{5/2}}
\nonumber \\
 &\simeq& 1 \times 10^{20} \frac{q|A|\sin\delta}{\lambda^{3/2}}
\left(\frac{m_{3/2}}{100 {\rm GeV}}\right)
\left(\frac{1 {\rm TeV}}{m_{\phi}}\right)^{5/2}.
\label{QballCharge}
\end{eqnarray}
Unless $Q>\mathcal{O}(10^{18})$, $Q$-balls will evaporate before 
decaying~\cite{BanerjeeJedamzik}. For $Q$ as in
Eq.~(\ref{QballCharge}), $Q$-ball decay temperature is $T_d \simeq1\gev$
to $1\mev$~\cite{Td,emd98-bbb}.  For example~\cite{Td},
\begin{eqnarray}
T_d\lsim 2\gev \times \left(\frac{0.03}{|K|}\right)^{1/2}
\left(\frac{m_\phi}{1\tev}\right)^{1/2} \left(\frac{10^{20}}{Q}\right)^{1/2} 
\nonumber
\end{eqnarray}
which is lower than the typical freeze out temperature of WIMPs,
$T_f\simeq \mchi/24$. Thus, the LSPs generated in $Q$-ball decays do not
subsequently thermalize. Nor will the baryon asymmetry be washed out
by sphaleron effects since $T_d<T_{ew}$~\cite{emd98-bbb}.  Note that
$Q$-balls decay prior to BBN and thus do not spoil its successful
predictions.

In the EMD scheme, $\treh$ also must be rather low. (This justifies
neglecting thermal effects in Eq.~(\ref{ADPotential}).) 
In fact, unless $\treh\lsim10^{3-5}\gev$, $Q$-balls could
thermalize~\cite{emd98-bbb}. 
In order to preserve the $\omegab$ -- $\Omega_{DM}$ relation,
one needs to suppress the neutralino
population from freeze-out. For this to happen it would be sufficient to
assume $\treh\lsim T_f$.

It is easy to see why in the EMD scenario, the ratio
$\omegab/\omegachi$ should be less than 1. 
The $Q$-ball is basically a huge ``bag'' of
squarks. It decays predominantly via $\squark\ra q+\chi$. Thus, for
one unit of a baryon number, at least $N_\chi \geq 3$ units of 
nonbaryonic number density are created.
(This number can be larger than 3 if one
takes into account additional decays of squarks into heavier charginos
and neutralinos which then cascade decay into the lightest
neutralino, which are model dependent.)
In other words, the LSP number density $n_\chi$ after $Q$-ball decay
is given by $n_\chi=N_\chi f_B n_b$,
where $f_B$ is the fraction of baryon asymmetry carried by the
AD field $\phi$ that is transferred into $Q$--balls. From
lattice calculations, $f_B\simeq 1$~\cite{QballKasuya}. 
Assuming that the LSPs subsequently do not undergo any
significant self-annihilation, and since in general 
$\Omega h^2=m Y= m\, n/s$,
this 
can be recast into
\begin{eqnarray}
\frac{\omegab}{\omegachi} = \frac{m_n Y_b}{\mchi Y_\chi}=
\frac{m_n}{\mchi} \frac{1}{f_B N_\chi},
\label{eq:nwimp-to-nb-abund-ratio}
\end{eqnarray}
where $m_n$ denotes the mass of a nucleon and $\mchi$ the mass of the
neutralino.  It is clear that Eq.~(\ref{eq:nwimp-to-nb-abund-ratio})
implies $\omegab/\omegachi$ to be less than one but not $\ll 1$. In the
EMD scenario, not only both types of matter are simultaneously produced
but also a right ratio of their abundances is predicted.

Unfortunately, this attractive picture runs into a serious problem of
over-producing neutralinos, as noticed already by EMD 
themselves~\cite{emd98-decayqballs,emd99-adcollapse}. 
Since $Y m \simeq 3.9 \times
10^{-10}\left(\abund/0.11\right) \gev $, one can
rewrite Eq.~(\ref{eq:nwimp-to-nb-abund-ratio}) as
\begin{eqnarray}
\mchi \simeq 1.5 \gev \left(\frac{3}{N_\chi}\right) \left(\frac{1}{f_B}\right)
\left(\frac{0.86\times 10^{-10}}{n_b/s}\right) \left(\frac{\abundchi}{0.11}\right).
\label{eq:mchibound}
\end{eqnarray}
In order to remain consistent with the values of $n_b$ and $\abundchi$
derived from observations, the neutralino mass has to be
$\mathcal{O}(1\gev)$ which, in the MSSM, is excluded by
LEP~\cite{pdg06}.  Here, we have neglected a possible contribution to
the LSP density from freeze-out.  If it were
significant, the problem would become only worse.
Moreover, the condition~(\ref{eq:mchibound}) puts into
question an attractive AD mechanism in a large class of
SUGRA models. 

To circumvent these problems, one has to review assumptions in the above discussions,
namely: 

(i) LSPs produced in $Q$-ball decay do not annihilate;

(ii) The LSP is the lightest neutralino of the MSSM.

If we relax assumption (i), the neutralino LSP with the mass of
$\mathcal{O}(10^2\gev)$, consistent with LEP, becomes acceptable. Indeed,
allowing for subsequent LSP self-annihilation, the LSP density will
be reduced by $\sim \langle\sigma_\chi v\rangle T_d$~\cite{FujiiHamaguchi}.
If the cross section $\sigma_\chi$ for the LSP (self-)annihilation is large
enough, e.g., when the LSP is Higgs-ino or
Wino-like~\cite{FujiiHamaguchi}, 
the relation between $\omegab$ and
$\Omega_{DM}$ is lost. One interesting exception is when
the energy density of universe is dominated by $Q$-ball
itself~\cite{fy02}.

Alternatively, if we lift assumption (ii), the
$\omegab$--$\Omega_{DM}$ relation may be preserved.  One way is to
consider, e.g., models with the Higgs sector supplemented by a singlet.
If its fermionic partner, the singlino, is the LSP then, for some
specific choices of parameters~\cite{Flores:1990bt}, it could be
possible to circumvent the LEP bound and perhaps also to suppress the
LSP abundance from freeze-out.  
In the rest of the paper, we will investigate axino LSP as DM.

\vspace{0.15cm}\noindent
{\bf 3. Axino dark matter from $Q$-balls.}\hspace*{0.3cm}

In general, axinos, like gravitinos, can be produced in both thermal
processes (TP) and in nonthermal processes (NTP), e.g., in late
decays.  
TP consists of the scatterings and the decays of particles in the thermal bath.  
NTP is given by the decay of
the Next-to-LSP (NLSP) relic (which, for simplicity, we assume to be
the neutralino) from freeze-out or from the decay of $Q$-balls in our
scenario.

The relevant Boltzmann equations can be written as
\begin{eqnarray}
&& \dot{n}_{\chi}+3Hn_{\chi} =
 -\langle \sigma_{\chi} v\rangle (n_{\chi}^2 - n^2_{\chi,{\rm eq}})
 +\gamma_Q  -\Gamma_{\chi}n_{\chi}, \\
&& \dot{n}_{\axino}+3Hn_{\axino} =
 \langle\sigma v\rangle{}_{ij} n_i n_j
 + \langle\sigma v\rangle{}_{i} n_i +\Gamma_{\chi}
n_{\chi},
\end{eqnarray}
where $\langle \sigma_{\chi} v\rangle$ is the usual neutralino
freeze-out term, $\gamma_Q$ denotes the contribution to $\chi$ from
$Q$-balls decay, $\Gamma_{\chi}$ is the decay
rate of the neutralino, 
$\langle\sigma v(i+j \rightarrow \axino+...)\rangle{}_{ij}$ and 
$\langle\sigma v(i \rightarrow \axino+ ...)\rangle{}_i$ 
are the scattering cross section and the decay rate
for the thermal production of axinos.
 
The total NLSP abundance is given by
\begin{eqnarray}
Y_{\chi} = N_{\chi} f_B\frac{n_b}{s} + Y_{\chi}^{TP}
\end{eqnarray}
where
$ Y_{\chi}^{TP} \simeq H/s_{|_{T=m_{\chi}}} \mchi/T_f/{\langle\sigma v\rangle_{ann}}$, 
as usual. Since $n_{\axino}= n_{\chi}$, owing to R-parity conservation, 
the resulting number density of axino is given by
\begin{equation}
Y_{\axino} = Y_{\axino}^{NTP}+ Y_{\axino}^{TP} ,
\end{equation}
with
\begin{equation}
Y_{\axino}^{NTP} =
 1 \times 10^{-10} N_{\chi}  \left(\frac{f_B}{1}\right)
 \left(\frac{n_b/s}{1\times 10^{-10}}\right) + Y_{\chi}^{TP},
\end{equation}
where $Y_{\axino}^{TP}$ denotes the axinos produced by thermal
processes.  Since typically $Y_{\chi}^{TP} \sim 10^{-11}$, one can see
that nonthermal production of axinos due to the thermal relic NLSPs
decay can easily be negligible compared to that from $Q$-ball
production, and its contribution to $\Omega_{\axino}h^2$ is further
suppressed by the ratio $\maxino/m_{\chi}$.
The thermally produced axino $Y_{\axino}^{TP}$ also
can be subdominant, say $Y_{\axino}^{TP} \lesssim 10^{-11}$, for
$\treh \lsim 100\gev$~\cite{ckkr,AxinoCMSSM}.
Hence, the axino dark matter density is estimated as
\begin{eqnarray}
 \left(\frac{\Omega_{\axino}h^2}{0.11}\right) \simeq 
\left( \frac{ \maxino} {1.5 \gev} \right) \left(\frac{ N_{\chi}
    f_B}{3}\right) \left( \frac{\abundb}{0.02}\right).
\end{eqnarray}
One can see that
the baryon asymmetry and the dark matter abundance are readily linked.

As mentioned above, axino mass is strongly model dependent; in
particular, it critically depends on how the visible and hidden sectors
are coupled~\cite{axinomass,axinomass-my,axinomass-gy}. 
At tree level, either $\maxino={\cal
O}(m_{3/2})$ or ${\cal O}(m_{3/2}^2/f_a)={\cal O}(\!\kev)$. However, 
in the latter case, trilinear terms can generate a substantial 1-loop
correction of order $f_Q^2/8\pi^2 A$, where $f_Q$ is the Yukawa
coupling of the heavy quark to a singlet field containing the axion,
which gives $\maxino$ in the range of a few tens of GeV or
less~\cite{axinomass-my,axinomass-gy}.

The final check point is the compatibility with successful predictions
of BBN.  However, this is not really a problem for the axino LSP
because its interactions are less suppressed than those of the
gravitino, roughly by $\left(M_P/f_a\right)^2$ and, so long as the
NLSP is heavier than about $150\gev$, axinos are produced before the
time of BBN~\cite{ckkr}. In contrast, the gravitino LSP would be
produced in late NLSP neutralino decays, which faces strong
constraints from BBN~\cite{Seto,rrc}.

\vspace{0.15cm}\noindent
{\bf 4. Conclusions.}\hspace*{0.3cm}

We have shown that the framework with cold dark matter axino LSP
produced in $Q$-ball decays can explain the abundance of dark matter and
the baryon asymmetry simultaneously and may be an answer to the
$\omegab \sim \Omega_{DM}$ coincidence.  In this approach, the
similarity between $\omegab$ and $\Omega_{DM}$ is explained by
basically only the axino mass of order of (sub-)GeV.
The essential property of $Q$-ball decays is that one can predict the
number of SUSY particles per one baryonic charge from $Q$-ball
$N_{\chi}$. 
A characteristic feature is low reheat temperature $\treh$ of $10^2$ GeV.

%

\vspace{0.15cm}\noindent {\bf Acknowledgments.}\hspace*{0.3cm} The
work of O.S. is supported by PPARC. We thank the European Network of
Theoretical Astroparticle Physics (ENTApP), part of ILIAS, under
contract number RII3-CT-2004-506222 and the EC 6th Framework Programme
MRTN-CT-2004-503369 for partial financial support.




\end{document}